# Cerebral Autosomal Dominant Arteriopathy with Subcortical Infarcts and Leukoencephalopathy (CADASIL): Immunotherapy and Cell Therapy approaches

Sierra Haile[1] (ORCID: 0009-0000-4649-7006), Benjamin C. Balzer[1] (ORCID: 0009-0009-5463-6999), Emily Egan[1] (ORCID: 0009-0005-8413-0172), Cheryl L. Jorcyk[1] (ORCID: 0000-0001-5715-472X), Nilufar Ali[1] (ORCID: 0000-0003-0010-0821),

[1] Department of Biological Sciences, Boise State University, Boise, ID, USA.

**Author Contributions:**
All authors contributed to the conceptualization of the study. BSU graduate students SH, EE, and BB conducted the literature review and wrote the initial draft of the manuscript. CJ and NA provided final revisions to the manuscript and offered supervision throughout the process.

**Corresponding Author(s):**
Nilufar Ali (nilufarali@boisestate.edu)

**Abstract**
This review article focuses on current and emerging therapeutics for CADASIL (Cerebral Autosomal Dominant Arteriopathy with Subcortical Infarcts and Leukoencephalopathy). CADASIL is an inherited vascular disease that impairs blood flow in the small cerebral vessels of the brain, leading to strokes and other neurological deficits. The disease is caused by a mutation in the NOTCH3 gene located on chromosome 19. NOTCH3 encodes a transmembrane receptor expressed on vascular smooth muscle cells. In CADASIL, mutations in the NOTCH3 gene lead to the accumulation and deposition of the receptor, affecting the number of cysteine residues in its extracellular domain. These mutations result in the loss or gain of a cysteine residue within the epidermal growth factor-like repeat (EGFr) domains of the NOTCH protein. Beyond traditional symptomatic treatments for stroke, this work highlights advances in disease-modifying approaches—including gene editing, cell therapies, and immune-based interventions—aimed at altering the course of CADASIL. It also examines ongoing clinical trials and recent patents related to these novel strategies. In addition to summarizing diagnostic methods and molecular mechanisms, the article emphasizes the translational potential of current research and the experimental models driving therapeutic development. The goal is to offer a comprehensive overview of CADASIL and emerging interventions that hold promise for improving long-term outcomes.

**Keywords:**
**CADASIL, Immunotherapy, Gene Therapy, Treatment, Nucleic Acid Therapy**

**Statements and Declarations:**
This study was funded by the following grants: BSF 2017237; NIH grants P20GM103408, P20GM109095, R25GM123927, U54GM104944-06, and 1C06RR020533; the METAvivor Quinn Davis Northwest Arkansas METSquerade Fund, the AR298 Smylie Family Cancer Fund, the M.J. Murdock Charitable Trust, and the Boise State University Biomolecular Research Center (BRC).

**Introduction:**

CADASIL was first identified in 1977 and classified as a hereditary multi-infarct dementia. CADASIL affects small arteries in the brain (cerebral arterioles), and is a hereditary disease [1,2]. CADASIL was mapped to chromosome 19q12 in 1993 and was linked to mutations in the NOTCH3 gene in 1996 [3,4]. There are at least 200 different pathogenic variants of NOTCH3 that have been identified [5]. These mutations tend to result in an odd number of cysteine residues and are associated with an accumulation of NOTCH3 protein in the cerebral arterioles [6,7]. The disease is not well characterized within the clinical medical community and has symptomatic heterogeneity for the disease, even within families [8]. Diagnosis often occurs by considering family history for migraines, strokes, and dementia alongside MRI findings of white matter changes and genetic testing [8]. Current treatment options for patients can manage the symptoms but not the underlying disease itself, although treatments to address the root cause of CADASIL are currently being researched [9].

**Table I: Common Clinical Findings in CADASIL Patients [5]**

| Country | France |
|---|---|
| Number of Patients | 446 |
| Number of Families | 298 |
| Sex Distribution: Male, n (%) | 197/246 (44) |
| Mean Age at Study | 53 ± 12 |
| Mean Age at Diagnosis | 51 ± 12 |
| Family History of Stroke, n (%) | 235 (53) |
| **Trait** | **Clinical Manifestation** |
| Hypertension, n (%) | 95 (28) |
| Hypercholesterolemia, n (%) | 168 (49) |
| Smoking (Current or Ex.), n (%) | 208 (61) |
| Diabetes, n (%) | 23 (6.7) |
| Symptomatic, n (%) | 412 (93.4) |
| Gait Ataxia, n (%) | 131 (30) |
| Urinary Incontinence, n (%) | 89 (20) |
| Pseudobulbar Palsy, n (%) | 19 (4) |
| Headache, n (%) | 229 (67.6) |
| Mean age at Onset | NA |
| First Manifestation, n (%) | 131 (29.7) |
| Migraine, n (%) | 221 (52) |
| Migraine with Aura, n (%) | 179 (40) |
| Stroke or TIA, n (%) | 235 (53.2) |
| Mean Age at Onset | 42 ± 10 |
| First Manifestation, n (%) | 158 (35.8) |
| Cognitive Impairment | 191 (43.4) |
| Dementia DSM IV, n (%) | 42 (9.5) |
| Mean Age at Diagnosis | 57 ± 11 |
| First Manifestation, n (%) | 26 (5.9) |
| Depression, n (%) | 162 (36.3) |
| Manic Episodes, n (%) | 4 (<1) |
| Delirium, n (%) | 1 (<1) |
| Schizophrenia, n (%) | NA |
| Epileptic Seizure, n (%) | 38 (8.6) |
| First Manifestation, n (%) | 10 (2.4) |

**Clinical Presentation:**

The clinical presentation of CADASIL is heterogeneous. In 1998, four major presentations were described: ischemic deficits (strokes or transient ischemic attacks), cognitive deficits (dementia), headaches (migraines with aura), and psychiatric disturbances (mood changes) [10]. In 2009 the forms of clinical presentation were expanded upon to include apathy [11]. The most recent clinical presentations are listed in **Table 1**, adapted from Dupé et al. 2023 [5]. This 446-person study showed that 412 patients (93%) were diagnosed with clinical manifestation of symptoms [5]. The most diagnosed symptom was categorized as headaches among 299 patients (67.6%). Stroke and transient ischemic attacks being the second most diagnosed symptom with 235 patients (53.2%). The next-most diagnosed symptom was forms of cognitive impairments with 191 participants (43.4%). The least diagnosed symptom categorically was psychiatric disturbance with 167 individuals (37.8%) [5].

**Genetics and Hereditariness**

CADASIL is an autosomal dominantly inherited angiopathy, most patients have an affected parent; de novo pathogenic variants are seemingly rare [12]. CADASIL is generally caused by heterozygous mutations, but homozygous mutations, atypical mutations, and null mutations have also been reported [13]. The majority of the mutations are mis-sense mutations that lead to an odd number of cysteine residues [14]. As there is limited research, it is hypothesized that the prevalence of CADASIL is estimated at between 0.8-5 in 100,000 but this disorder is often under-diagnosed, or misdiagnosed [15]. Neuroanatomically, male patients with CADASIL experienced greater brain atrophy in the bilateral frontotemporal cortex, including the orbitofrontal cortex, ACC, entorhinal cortex, and right temporo-occipital regions [16].

**Mechanism of Disease**
NOTCH receptors encompassing NOTCH1 through 4, are integral transmembrane proteins that are pivotal in dictating cell differentiation and function. All NOTCH receptors contain an intra and extra cellular domain, and after binding to a transmembrane ligand the extra-cellular domain (ECD) is cleaved [17]. Their activation is canonically triggered by five different ligands belonging to the Jagged and Delta-like families. The cleavage of the ECD initiates a cleavage by y-secretase that releases the NOTCH intracellular domain (ICD) which functions as a transcription factor [17]. NOTCH3 plays a crucial role in cell signaling pathways that maintain the stability and survival of vascular smooth muscle cells (VSMCs) [17]. Induced pluripotent stem cells that were differentiated into VSMCs derived from CADASIL patients displayed increased proliferation and cytoskeleton disorganization [18]. The arteriopathy appears to be systematic, however large effects are seen in cerebral small to medium-sized penetrating arteries and leptomeningeal arteries [19]. Other pathological changes of patients with CADASIL include cystic softening in the white matter, corpus callosum, internal capsule, basal ganglia, thalamus, and brain stem. A diffuse myelin loss in the hemispheric white matter and widespread neuronal apoptosis in the cerebral cortex has also been documented [19].

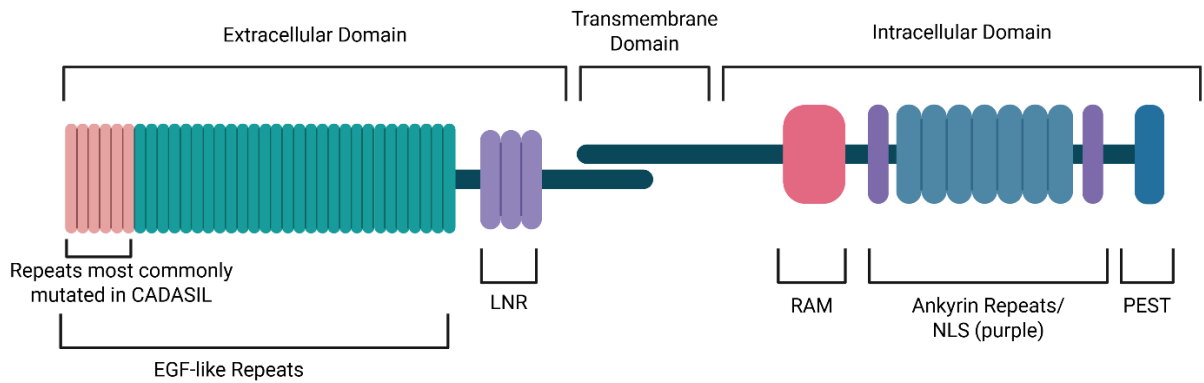

***Fig. I*** *NOTCH3 Receptor Intra and Extracellular Domains. The EGF-like repeats most frequently mutated in CADASIL are labeled and indicated in pink in the extracellular domain. The mutations typically lead to the addition or removal of a cysteine residue. The Notch/LIN-12 repeats (LNR) are not implicated in CADASIL. The intracellular domain functions as a transcription factor, made up of the RAM domain, Ankyrin repeats flanked by the nuclear localization sequences (NLS) and the proline-, glutamate-, serine- and threonine-rich domain (PEST) at the C-terminal end of the protein.*

NOTCH3 mutations are most frequently found in the outermost EGFr, these mutations typically lead to an odd number of cysteine residues and NOTCH3 aggregation, visualized in **Figure I** [5]. There are 36 EGF repeats in the ECD, as well as three cysteine-rich Notch/LIN-12 repeats (LNR) which are not implicated in the disease [17]. These unpaired cysteine residues are thought to form protein aggregates and have been shown to bind other transmembrane proteins [20]. These protein aggregates may alter extracellular matrix homeostasis, resulting in the fibrosis of small vessels in the brain, potentially leading to symptoms associated with CADASIL [6,20,21]. While a lack of NOTCH3 signaling may lead to CADASIL symptoms, this hypothesis has been questioned. NOTCH3 knock-out mice do not present characteristic CADASIL symptoms such as Granular Osmiophilic Material (GOM) deposits and the unique small vessel pathology of CADASIL [20,22]. However, the NOTCH3 knock-out mice do present some symptoms consistent with CADASIL, such as VSMC loss, disorientation, and morphological change [22].

The NOTCH3$^{ECD}$ has been shown to aggregate in GOM deposits, which are the hallmark characteristic of CADASIL [23,24]. These granular deposits show up as dark aggregates in electron microscopy and were first identified in a case study published in 1993 [25]. GOM deposits are present in arterioles throughout the body and are typically located in VSMCs [23]. Immunohistochemical analysis of dermal tissue samples show that an accumulation the mutated NOTCH3$^{ECD}$ is consistent with the distribution of GOMs, indicating that the pathogenic NOTCH3 mutations may contribute to the formation of GOMS and disease [23]. Dermal tissue samples are typically used to diagnose CADASIL because the GOM deposits are well preserved and the tissue samples are relatively easy to collect [26].

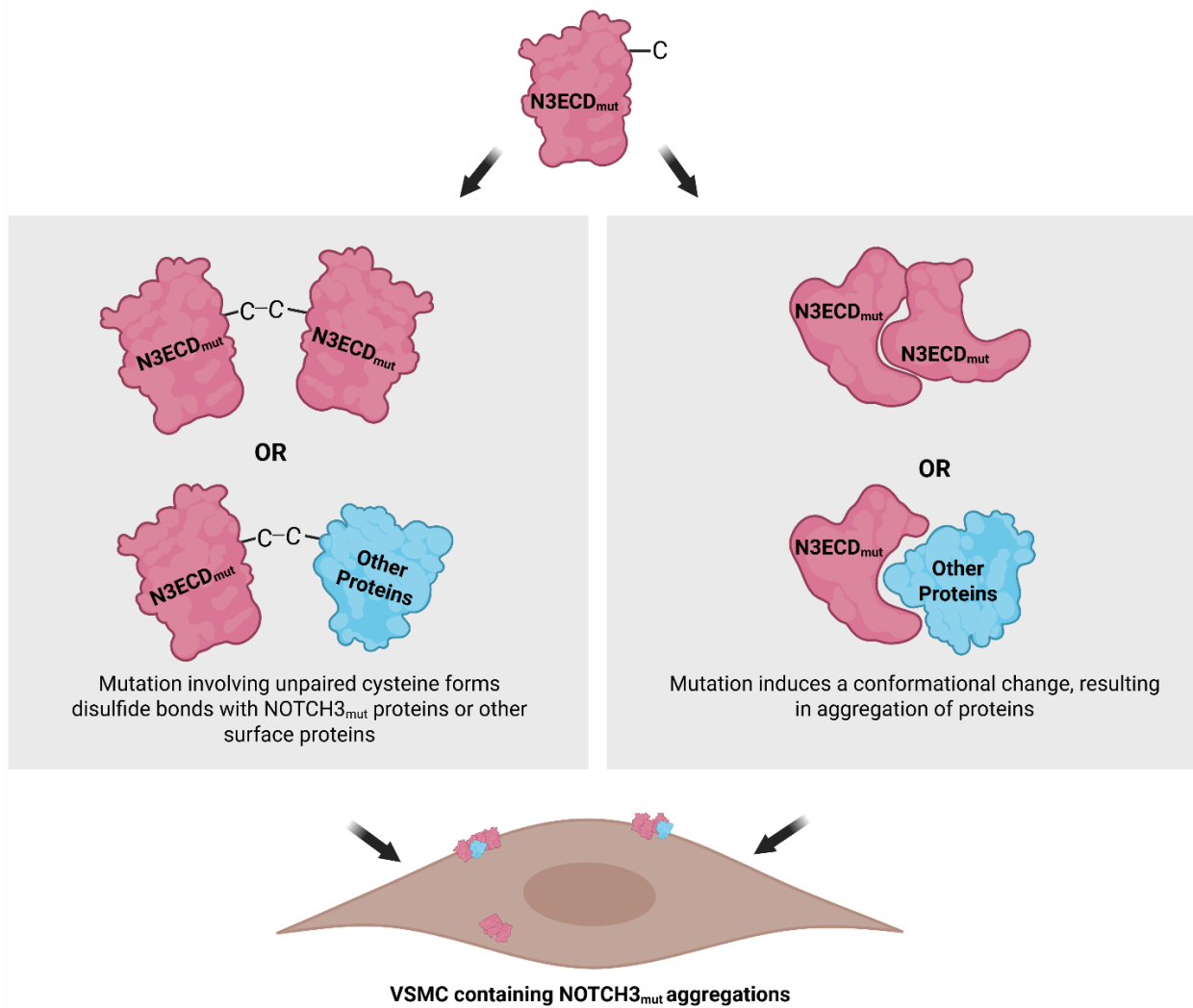

***Fig. II*** *Potential mechanisms of protein aggregation in CADASIL. Protein structures depicted do not reflect the actual conformations of NOTCH3 mutations. NOTCH3$^{ECD}$ (N3ECD) in CADASIL is thought to form self-aggregates or aggregate with other cell surface proteins. The unpaired cysteine residues may be forming disulfide bonds with other NOTCH3 proteins or non-NOTCH3 cell surface proteins. This change in cysteine residues could also lead to a conformational change, promoting binding of mutated NOTCH3 to other mutant NOTCH3 proteins/non-NOTCH3 proteins. These NOTCH3 aggregates appear in GOM deposits and have been identified in the endoplasmic reticulum of HEK 293 cells as well.*

Many mechanisms for the pathophysiology of CADASIL have been proposed, multiple of which may be true as different mutations could result in different aggregation mechanisms. With many of the mutations having an unpaired cysteine, it's possible that the cysteine residues may be forming disulfide bonds with other proteins, or this change in cysteine residues changes the tertiary structure of the protein which could allow for protein aggregation. These protein aggregations may be other mutated NOTCH3 proteins, wild-type NOTCH3, or other surface proteins [6,20,27,28]. These hypotheses are presented as a schematic in **Figure II**. Stress responses from the endoplasmic reticulum could play a role in the pathogenesis of CADASIL.

Two pathogenic NOTCH3 variants in HEK 293 cells were shown to colocalize and immunoprecipitated with calnexin and resist degradation [2]. The cell viability of the mutated NOTCH3 HEK 293 cells was also much lower, indicating that there may be an apoptotic response from the cells as a result of the aggregations. This could explain some of the disease pathology symptoms seen in CADASIL, such as VSMC degeneration. In another study using VSMC lines, some intracellular mutant NOTCH3 aggregates were seen, but a similar amount of aggregates were seen in wild-type NOTCH3 cells [29]. Depending on the mutation, O-linked glycosylations may be altered in mutated NOTCH3 proteins during processing in the cell, which may induce a conformational change in the protein and result in protein aggregation [30,31].

It is still unclear how GOM deposits result in ischemic stroke. Although anticoagulants are prescribed to 13.9%-48.2% of CADASIL patients who have experienced stroke, evidence supporting thrombotic origin of strokes in CADASIL has not been demonstrated [32–34]. It may be possible that the decreased lumen dimeter of the vessels may be a factor in the development of stroke. The lumen of white matter arterioles in elderly CADASIL patients was 1.8 times smaller than similarly aged control patients, and the vessel walls were found to be considerably thicker [23]. The mechanisms of other aspects of the disease are poorly defined, but a disrupted blood brain barrier may be a contributing factor. A breakdown in the blood-brain barrier has been linked to both cognitive impairment and may play a role in migraine which are both common factors of CADASIL [35,36].

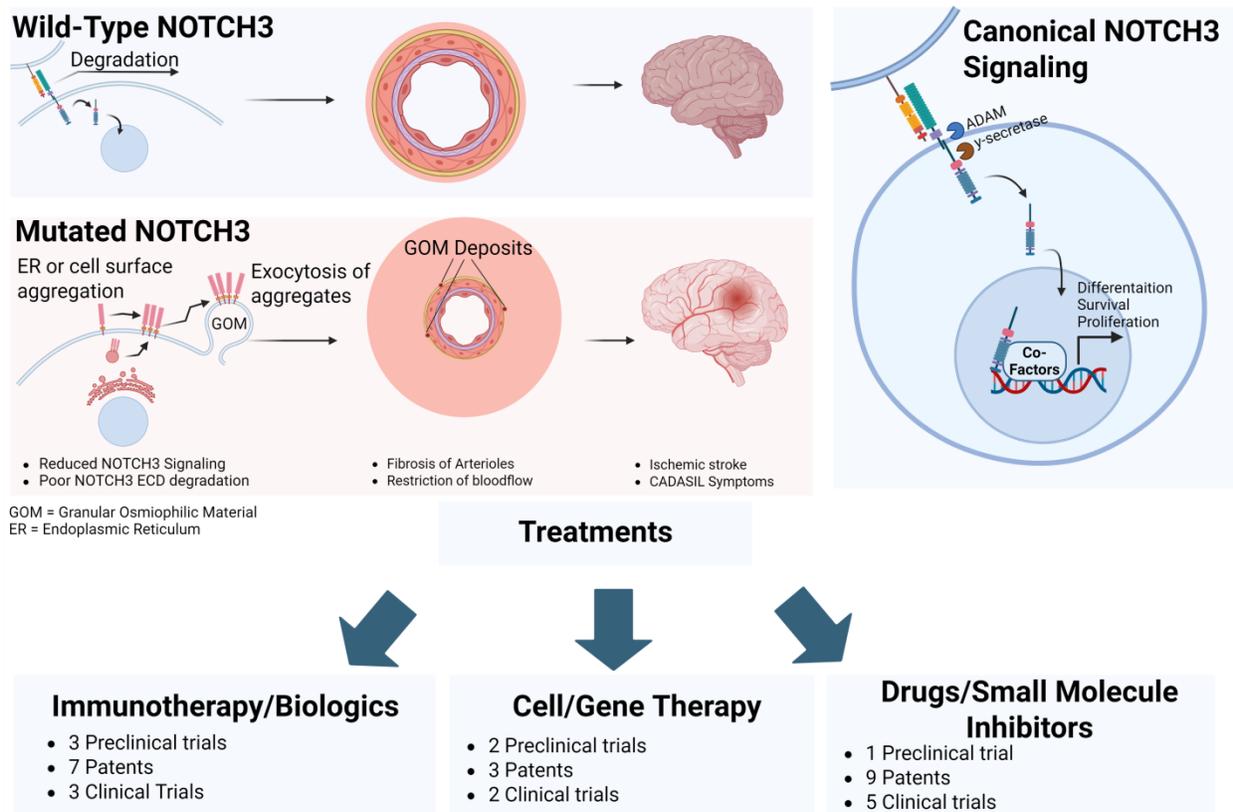

***Fig. III*** *Proposed mechanism of pathogenesis in CADASIL as a result of NOTCH3 mutation and possible therapeutics. In unaffected individuals the NOTCH3$^{ECD}$ will be degraded after ligand*

*binding (typically from a neighboring cell), followed by two consecutive protease cleavages via ADAM proteases and y-secretase. This results in the release of the intracellular domain which functions as a transcription factor. In CADASIL, the mutated NOTCH3$^{ECD}$ aggregates on the cell surface or the endoplasmic reticulum (ER) and NOTCH3 signaling is reduced, presumably leading to small scale changes in the arterioles which ultimately result in ischemic stroke and other CADASIL symptoms. Current therapies to treat CADASIL include immunotherapies, biologics, cell and gene therapies, and drugs/small molecule inhibitors.*

### Current Immunological, Cell, and Gene Therapies

Immunotherapy, cell and gene therapy are disease modifying treatment regiments that ultimately work to fight and eliminate a disease [37–45]. Each treatment investigates the genetic and molecular mechanisms of the immune system's response to disease and develops biological interventions to improve the health of diseased individuals. Immunotherapy works by activating or suppressing the immune system to better fight disease. The goal of immunotherapy is to enhance the immune system's ability to recognize, target, and eliminate disease-causing cells. Cell therapy introduces healthy cells into the body to replace damaged or diseased cells to better modulate the function of a patient's cells. Cell therapies work by the expression of factors, or direct interaction, or by the removal of the damaged or diseased cells by utilizing immune cells. Gene therapy is a treatment that alters diseased gene/s inside the body through its replacement, or by adding a new gene in an attempt to improve the body's ability to fight disease. Gene therapy aims to correct defective genes by replacing or fixing the mutated genes and making diseased cells more evident to the immune system. The hypothesized mechanism and a summary of the existing therapies, patents, and clinical trials are visually represented in **Figure III**.

### Immunotherapy in CADASIL

NOTCH3 accumulations are a hallmark of the disease, thus immunotherapy targeting NOTCH3 aggregation or aggregates may be an attractive therapeutic strategy. CADASIL shares some important features with Alzheimer's disease, as both are characterized by protein extracellular aggregations. Thus, it is conceivable that NOTCH3 aggregates are accessible to efficient antibody-mediated clearance, as has been demonstrated for amyloid-beta immunotherapies [38].

A preclinical active immunization therapy targets aggregated NOTCH3$^{ECD}$ around VSMCs in a CADASIL mouse model (TgN3R182C). The mice express a commonly mutated human NOTCH3 receptor which allows the development of progressive cerebrovascular NOTCH3$^{ECD}$ aggregations and GOM deposits, both of which are pathological hallmark characteristics seen in people with CADASIL. The mice were immunized with protein aggregates of EGF domains 1-5 that contained the R113C mutation. The proposed method included the generation of an antigen that is selective for aggregated NOTCH3 and spare monomeric NOTCH3 receptors, which would minimize adverse effects on NOTCH3 signaling. The mutation in the mouse model was intentionally different than the immunogen produced to demonstrate that this treatment may be effective in treating multiple variants of NOTCH3. Results demonstrated that mice repeatedly immunized with the aggregated mutant NOTCH3 peptide displayed a 38-48% reduction of $NOTCH3^{EDC}$ deposition around the brain capillaries, and reduced levels of the $NOTCH3^{ECD}$ in the blood-levels. This form of active immunization did not disrupt normal NOTCH signaling nor cause any kidney toxicity, inflammation, or neurodegeneration of VSMCs in the vasculature of the immunized mice [38].

Monoclonal antibodies (mAb) also present a promising immunotherapy option for CADASIL, similar to their application in neurodegenerative diseases [46]. Ghezali et al. tested a passive immunotherapy for CADASIL targeting the $NOTCH3^{ECD}$ with a mAb in the TgN3R169C mouse model [37]. The 5E1 mouse mAb was designed against the EGF repeats (17-21) of human NOTCH3, this antibody strongly co-localizes to vascular $NOTCH3^{ECD}$ deposits when peripherally administered, measured by immunohistochemistry [37]. The 5E1 mAb treatment did not affect the number or surface area of $NOTCH3^{ECD}$ deposits, it also did not promote perivascular microglial activation and attenuation of $Notch3^{ECD}$ or GOM deposition [37]. In mice the 5E1 mAb treatment improved the blood flow in mice comparable that to the wild type population and had no effect on white matter lesions [37].

These results indicate that there may be additional challenges using a passive immunization for CADASIL. Ghazali et al. proposes that microglial activation may be reduced in CADASIL mice or that passive clearance of the aggregates through the vessels in the brain is poor. A concern of antibodies being able to cross the blood-brain barrier effectively was also mentioned [37]. Both Alzheimer's disease and CADASIL are classified as protein elimination failure angiopathy (PEFA) disorders. Several reviews have claimed that CADASIL symptoms may be exacerbated by a lack of drainage of the protein aggregates, although it is unclear if a lack of interstitial fluid (ISF) drainage is actually a factor in the removal of protein aggregates in CADASIL [47–49].

Machucha-Parra et al. found that an NOTCH3 agonist antibody, A13, induced activation of the NOTCH3 signaling pathway in cell culture and transgenic mouse models regardless of mutations in the $NOTCH3^{ECD}$. This was evaluated using a endostain/collagen $18\alpha1$ system was used as a way to observe modulation of NOTCH3 activity [50]. The transgenic mice expressed the human C455R mutation of NOTCH3, this model was selected because it is associated with early onset of disease and the lack of signaling is believed to contribute to these symptoms [50]. These results show that the NOTCH3 pathway can be rescued in mice with a pathogenic CADASIL mutation using agonist antibodies which was shown to reduce mural cells death in arterioles. This study does not access the GOM deposits, evidence of stroke, or other manifestations of CADASIL post-treatment.

**CADASIL as a Candidate for Cell and Gene Therapy**

There are many aspects of CADASIL's disease pathology that make it a good target for gene therapy. The first is that according to current research, NOTCH3 mutations and aggregations are likely responsible for the downstream events that lead to CADASIL symptoms [20,22]. Secondly, the disease-causing protein is highly differentially expressed in smooth muscle cells, ideally minimizing potential off-target effects [51]. This disease is an excellent target for gene therapy as it appears to be monogenic. This is an especially appealing disease since many of the disease causing variants are single-nucleotide polymorphisms (SNPs) and should be able to be repaired using CRISPR technology, without the induction of a double-stranded break by Cas-9 [9]. Despite its promise the potential challenges of gene therapy in CADASIL could be the delivery of drugs through the blood-brain barrier, and that therapy would not repair existing infarctions. For an immunotherapy to be successful it would also require early diagnosis of CADASIL to prevent infarction and other permanent symptoms associated with the disease.

Gene therapy is a rapidly advancing field with various approaches for treating genetic diseases. Gene therapy approaches can be broadly classified as ex vivo and in vivo. Ex vivo gene therapy involves extracting autologous patient's cells, typically from blood or skin, and genetically modifying them in a laboratory using an expression vector to introduce a healthy gene [52]. These modified cells are then reintroduced into the patient's body. This method allows for precise targeting of specific cell types relevant to the disease and can potentially provide long-lasting effects as the modified cells can continually produce the therapeutic protein. However, it is a complex, multi-step, prolonged and invasive process consisting of a risk of immune rejection as the body might recognize the modified cells as foreign [52]. Additionally, not all cell types are easily manipulated and reintroduced effectively.

In vivo gene therapy, a newer and less invasive approach, bypasses the need to modify patient cells outside the body [52]. Instead, the gene therapy vector is delivered directly into the bloodstream or targeted tissues. The vector then enters cells within the target tissue, ideally delivering the therapeutic gene into the cell's nucleus where it starts producing the desired protein product. There are two main types: viral and non-viral vectors. Viral vectors use modified viruses to deliver the gene and are highly efficient but raise safety concerns due to potential viral activation or immune response. Non-viral vectors, which are not derived from viruses, are generally considered safer but can be less efficient in delivering the gene compared to viral vectors. In vivo gene therapy is simpler and has potentially broader applications, as it can be used for a wider range of diseases affecting various tissues. Furthermore, it reduces the risk of immune rejection since patient cells are not modified. However, challenges remain in ensuring the vector reaches the target cells effectively, navigates into the cell's nucleus, and controls the dosage to deliver the right amount of therapeutic gene. The immune system may still recognize the vector as foreign and attempt to clear it, presenting an additional challenge.

Jalil et al. utilized a stem cell reprogramming approach to modify mutated genes with the goal of treating monogenic diseases [53]. This method uses primary fibroblasts from the patient skin biopsies, which are then modified using a CRISPR-Cas9 adenine base editor to one of the SNPs responsible for CADASIL (475C>T) [53]. According to Opherk et al., this mutation represents 14% of the 371 individuals who had a CADASIL diagnosis with a mutated NOTCH3 gene [9]. Out of this population, 98.6% of the individuals had a SNP and may have the potential to be repaired by adenine or cytosine base editors, assuming the mutations are accessible to a CRISPR-Cas9 system [9]. While the reprogrammed fibroblasts were shown to have been successfully edited to carry the non-pathogenic variant, a change in phenotype was not demonstrated in the paper [53]. It is also unclear whether the modified fibroblasts would change the disease presentation, as the NOTCH3 mutation typically most strongly affects VSMCs. A similar injectable CRISPR-Cas9 system could also be used to target the most heavily affected cells as well. A more conclusive in vivo model would need to be tested before considering this therapy as a viable way to treat CADASIL.

A challenge that may come along with using cell therapy to treat CADASIL is the integration of edited VSMCs into the cerebral vessels. Based on existing research, the disease mechanisms seem to be due to an expression of mutated NOTCH3 by VSMCs, not a dysregulation of fibroblasts leading to fibrosis. This indicates that edited VSMCs would be needed to resolve disease symptoms, such as reduced cell proliferation [2,23]. The mechanism of stenosis in the arterioles of CADASIL patients is not entirely understood, but fibrosis of the vessel walls does appear to be a common symptom [23]. Whether these symptoms can be attributed to fibroblast dysregulation which could be resolved through gene editing is unknown, although edited fibroblasts may not resolve VSMC cell death seen in CADASIL models and patient pathology [23,54]. A strategy that could be effective would be using a viral vector to deliver a Cas-9 construct to the cells. Mice modeling hypertrophic cardiomyopathy were treated with an AAV9 Cas-9 base editing construct and had reduced fibrosis in the heart tissue [55]. Utilizing a Cas-9 base editor construct in a viral vector to edit VSMCs in CADASIL is a treatment strategy that should be considered [3]. Thus, while no approved gene therapies for CADASIL exist yet, researchers are actively exploring potential strategies, including developing vectors that can efficiently deliver gene therapy to brain microvasculature, refining gene editing techniques to precisely correct the NOTCH3 gene mutation, and conducting preclinical testing in animal models to assess safety and efficacy before human trials.

**Antisense Oligonucleotides and Exon Skipping**

Antisense oligonucleotide therapy is another approach that may be particularly effective in the treatment of CADASIL, as silencing the mutated gene or removing problematic exons could theoretically halt progression of disease. These types of therapies are currently being explored for the prevention and treatment of ischemic stroke [56,57]. Most ischemic strokes are multifactorial, environmental and genetic factors both contributing to their development [58,59]. There is a small subset of ischemic strokes, as seen in CADASIL, that have monogenic causes. While mitigating typical risk factors such as hypertension and smoking may help to prevent stroke in CADASIL patients, this recommendation has been informed by observational studies but has yet to be studied in a randomized trial [60]. Other risk factors have yet to be implicated in CADASIL and nucleotide based therapies could help to reduce the risk of stroke that cannot be reduced with existing therapeutics or lifestyle changes [60]. It has been shown that mutations in the NOTCH3 gene lead to cell surface aggregates, however the intracellular cytoskeleton is also affected. The vascular smooth muscles are not able to contract properly, causing aberrance in the auto regulation of small vessels in the nervous system [56].

Currently, there are no nucleotide based therapies that delay or alleviate the symptoms in CADASIL patients. It is hypothesized that the exclusion of the mutant EGFr domain from NOTCH3 would terminate the detrimental effect of the unpaired cysteine, preventing the toxic NOTCH3 accumulations and negative cascade events accompanied by the disease. An in-vitro proof of concept study tested this idea using pre-mRNA antisense mediated skipping of pathogenic NOTCH3 exons [61]. The antisense oligonucleotides which were tested targeted exons 2-3, 4-5, and 6. Patient-derived VSMCs, derived from CADASIL patients were transfected with the antisense oligonucleotides and the expressed proteins successfully skipped the targeted exons while maintaining functionality of the NOTCH3 protein, as determined by PCR analysis of downstream target genes. This novel application of exon skipping is a promising application in the development of rational therapeutic methods for up to 94% of CADASIL-causing mutations, as the vast majority of NOTCH3 mutations in CADASIL appear in exons 2-6 [5,61].

**Aggregation Targeting Therapies**

Aggregation-targeting therapies do not directly use the immune system to reduce disease symptoms, but target the formation of the protein aggregates on the VSMCs. DAPH (4,5-dianilinophthalimide) is a small molecule that was found to increase the rate of degradation of mutated NOTCH3 in edited HEK 293 cells [54]. These cells had tetracycline inducible mutant NOTCH3 expression, and when treated with DAPH, the presence of NOTCH3 in the cells was reduced [54]. These results were analyzed by western blot and immunohistochemistry. In this study, it's unclear if DAPH is also promoting the degradation of functional NOTCH3, as the HEK 293 cells degraded wild-type NOTCH3 by the experimental timepoint on their own after the tetracycline-induced expression [54]. The mechanism of action proposed in the paper is that the DAPH may be binding to a structural motif in the NOTCH3 aggregates; this is supported by the fact that the two mutations used had different levels of degradation [54]. This could indicate that the use of this small molecule may have varying success depending on the type of mutation being treated. Further research with this drug in vivo could show if the degradation of NOTCH3 is significant enough to reduce disease symptoms, as well as if the potential lack of signaling by the wild-type NOTCH3 expressed in heterozygous patients.

**Table II: Clinical Trials Focusing on the Diagnosis or Treatment of CADASIL**

| NCT Number | Study Title | Intervention | Phase | Status |
|---|---|---|---|---|
| NCT01865604 [a] | Impact of tDCS on Cerebral Autoregulation | DEVICE: Anodal tDCS<br>DEVICE: Cathodal tDCS<br>DEVICE: sham tDCS | NA | Completed |
| NCT04165213 [a] | Care of Persons With Dementia in Their Environments (COPE) in Programs of All-Inclusive Care of the Elderly (PACE) | BEHAVIORAL: COPE online training | NA | Completed |
| NCT02795052 [a] | Neurologic Stem Cell Treatment Study | PROCEDURE: Intravenous and Intranasal BMSC | NA | Recruiting |
| NCT04828434 [a] | Virtual Individual Cognitive Stimulation Therapy: a Proof of Concept Study | OTHER: Virtual Individual Cognitive Stimulation Therapy | NA | Active, Not Recruiting |
| NCT02267057 [a] | Efficacy of Pain Treatment on Depression in Patients With Dementia | DRUG: Paracetamol<br>DRUG: Buprenorphine<br>DRUG: Paracetamol placebo<br>DRUG: Buprenorphine placebo | PHASE4 | Completed |
| NCT03724136 [a] | Alzheimer's Autism and Cognitive Impairment Stem Cell Treatment Study | PROCEDURE: Intravenous Bone Marrow Stem Cell (BMSC) Fraction<br>PROCEDURE: Intranasal Topical Bone Marrow Stem Cell (BMSC) Fraction<br>PROCEDURE: Near Infrared Light | NA | Enrolling by Invitation |
| NCT04428112 [a] | Rural Dementia Caregiver Project | BEHAVIORAL: Building Better Caregivers Workshop<br>BEHAVIORAL: Attention Control | NA | Completed |
| NCT01361763 [a] | Safety Study of Dabigatran in CADASIL | DRUG: Dabigatran<br>DRUG: Antiplatelets | PHASE2 | Unknown |
| NCT00103948 [a] | The Efficacy, Safety, And Tolerability Of Donepezil | DRUG: Aricept | PHASE2 | Completed |

| NCT Number | Study Title | Intervention | | Status |
|---|---|---|---|---|
| | HCl (E2020) In Patients With CADASIL Who Have Cognitive Impairment | | | |
| NCT06072118 [a] | Adrenomedullin for CADASIL | DRUG: Adrenomedullin | | PHASE2 Completed |
| NCT04658823 [a] | Efficacy and Safety of Tocotrienols in CADASIL | DRUG: HOV-12020 (Palm tocotrienols complex) DRUG: Placebo | | PHASE2 Unknown |
| NCT04334408 [a] | Safety and Efficacy of Fremanezumab for Migraine in Adult CADASIL | DRUG: Fremanezumab DRUG: Placebo | | PHASE2 Withdrawn |
| NCT04753970 [a] | Retina is a Marker for Cerebrovascular Heath | DRUG: Cilostazol | | PHASE1 PHASE2 Recruiting |
| NCT05755997 [a] | CERebrolysin In CADASIL | DRUG: Cerebrolysin DRUG: 0.9 % NaCl | | PHASE2 Active, Not Recruiting |
| **NCT Number** | **Study Title** | **Intervention** | | **Status** |
| NCT05491980 [b] | Florida Cerebrovascular Disease Biorepository and Genomics Center | NA | | Recruiting |
| NCT02699190 [b] | LeukoSEQ: Whole Genome Sequencing as a First-Line Diagnostic Tool for Leukodystrophies | NA | | Completed |
| NCT03047369 [b] | The Myelin Disorders Biorepository Project | NA | | Recruiting |
| NCT05734378 [b] | Prognosis of Cerebral Small Vessel Disease | NA | | Recruiting |
| NCT02821780 [b] | CADASIL Disease Discovery | NA | | Completed |
| NCT05677880 [b] | Cerebral Autosomal Dominant Arteriopathy With Subcortical Infarcts and Leukoencephalopathy (CADASIL) Study | OTHER: Study Procedures | | Recruiting |
| NCT05567744 [b] | Registry for CADASIL | OTHER: Registry | | Recruiting |
| NCT05781139 [b] | Comparative Study Between Alzheimer's and Multi-infarct Dementia | DIAGNOSTIC TEST: levels of Neurofilaments (NfL) in serum | | Recruiting |
| NCT05793424 [b] | Establishment of a CSF Bank for the Development of Biomarkers of Smooth Muscle Cell (SMC) Damage in Monogenic Cerebral Small Vessel Disease | OTHER: Cerebrospinal fluid (CSF) sample and additional blood samples | | Unknown |
| NCT02071784 [b] | Imaging Study of Neurovascular Coupling in Cerebral Autosomal Dominant Arteriopathy With Subcortical Infarcts and Leukoencephalopathy (CADASIL) | NA | | Completed |
| NCT04036084 [b] | Development of New Biomarkers With Magnetic Resonance Imaging for Longitudinal Studies in CADASIL Angiopathy | DIAGNOSTIC TEST: Cerebral Magnetic resonance imaging (MRI) | | Unknown |
| NCT05473637 [b] | Taiwan Associated Genetic and Nongenetic Small Vessel Disease | DIAGNOSTIC TEST: MRI | | Recruiting |
| NCT01114815 [b] | Research Study on Cerebral Autosomal Dominant Arteriopathy With Subcortical Infarcts and Leukoencephalopathy (CADASIL) | NA | | Completed |
| NCT02032225 [b] | Generation of a Cellular Model of CADASIL From Skin Fibroblasts | OTHER: Skin biopsy | | Completed |
| NCT05902039 [b] | MRI Study of Blood-brain Barrier Function in CADASIL | OTHER: MRI | | Recruiting |
| NCT05072483 [b] | Natural History Study of CADASIL | NA | | Recruiting |
| NCT04310098 [b] | CADASIL Registry Study | NA | | Recruiting |
| NCT06148051 [b] | AusCADASIL: An Australian Cohort of CADASIL | NA | | Recruiting |
| NCT02837354 [b] | The Silent Cortical Infarcts in the Cerebral Amyloid Angiopathy: Is There a Link With Subarachnoid Hemorrhage? | OTHER: No intervention | | Completed |
| NCT06859658 [b] | Development and Validation of a Functional MRI Biomarker of Cerebral Small Vessel Dysfunction in CADASIL (fMRI BioSVD) | Radiation: Functional MRI at 3T | | Not Yet Recruiting |
| NCT06938100 [b] | Genotype, Clinical Features and Imaging of Neuroradiological Abnormalities in CADASIL (GENICa) | DIAGNOSTIC TEST: MRI OTHER: Skin biopsy | | Recruiting |
| NCT06933212 [b] | Effect of the Mediterranean Diet in Patients Affected by CADASIL and Cerebral Amyloid Angiopathy. (DIETETICA) | Other: Mediterranean Diet with Extra Virgin Olive Oil Other: Mediterranean Diet with Walnuts Other: Low-Fat Control Diet | | Recruiting |
| NCT06935578 [b] | RAre, But Not aLone: a Large Italian Network to Empower the Impervious diaGNostic Pathway of Rare cerEbrovascular Diseases (ALIGNED) | OTHER: Registry | | Recruiting |

<sup>a</sup> - Interventional ; <sup>b</sup>- Observational

**Clinical Trials**

As of June 2025, there are 37 clinical trials listed on clinicaltrials.gov for CADASIL broken down here into two categories of being interventional or observational **(Table II).** There are 14 interventional clinical trials and 23 observational trials for CADASIL therapies. Within the interventional trials, 5 are currently recruiting, with the rest being completed. There is a range of treatments being tested. One trial, NCT04165213, focuses on the Care of Persons with dementia in their Environments (COPE) training module to better elderly care.  Two trials, NCT02795052 and NCT03724136, focus on Intravenous and Intranasal Bone Marrow Stem Cell (BMSC) treatments (Table 2a.). Drug related trials such as NCT04753970 and NCT05755997 aim to act as immunotherapies for CADASIL Treatment. NCT04753970 is investigating Cilostazol, which is a quinolone derivative used to treat peripheral vascular disease (as approved by the FDA). It has been indicated for secondary prevention of TIA attacks or other forms of stroke [62]. NCT05755997 investigates Cerebrolysin, an enzymatic cocktail derived from pig brain, revolving around neurotrophic factors [63]. The trials not listed here explicitly are either complete, withdrawn, or of an unknown status. It is important to evaluate the current stages of development for CADASIL treatment in Interventional studies. All CADASIL specific studies are currently Phase 1 or Phase 2 in clinical trials. There is a long way to go before there are publicly available treatments for the disease. The treatment of CADASIL is still in a developmental stage.

The observational clinical trials for CADASIL are looking for individuals with the disease to contribute to the understanding of CADASIL through study. Databases formed with the intent of studying CADASIL or multiple different forms of vascular diseases. These databases are then used to inquire about people who may wish to join other studies. Diagnostic studies are also being developed, with examples being the use of MRI to better understand the brain matter of those afflicted with CADASIL. Other sampling tests, such as skin biopsies and blood samples are also being investigated. CADASIL Related studies suffer from the rarity of the disease within the general population. Due to the lower number of people the longer it takes to test and produce various forms of therapy or treatment.

**Table III: Patented Treatments, Diagnostic Tests, and Models for CADASIL**

| THERAPIES | | | | | |
|---|---|---|---|---|---|
| Patent Number | Title | Author/Company | Date Published | Symptom/ Disease Target | Method |
| UA129400C2 | Use of cerebrolysin to reduce mortality in patients with CADASIL | Штефан Вінтер, Штефан ВИНТЕР, Герберт Мьослер | 4/16/2025 | General CADASIL Symptoms | Cerebrolysin [f] |
| WO2025048409A1 | Pharmaceutical composition comprising htra activator as active ingredient for preventing or treating degenerative diseases, and method for screening htra activity modulator | 강민지,이지연,이지유 | 3/6/2025 | NOTCH3 aggregates via activation of HRTA | Activation of HTRA via small molecule [c] |
| WO2024243 | Method of treating patients with | Thomas Macallister, et | 11/28/2024 | General | Fasudil or |

| Patent Number | Title | Author/Company | Date Published | Symptom/Target | Method |
|---|---|---|---|---|---|
| 197A2 | NOTCH3 mutations | al. | | CADASIL Symptoms | hydroxyfasudil [c] |
| US20240269234A1 | Method of treating cerebral autosomal dominant arteriopathy with subcortical infarct and leukoencephalopathy (CADASIL) | Li-Ru Zhao, et al/Research Foundation of State University of New York | 8/15/2024 | General CADASIL Symptoms | Multi-growth factor supplementation [e] |
| JP7467626B2 | Use of compositions containing vitamin E tocotrienol for managing cerebral autosomal dominant arteriopathy with subcortical infarction and leukoencephalopathy (CADASIL) | ユエン, et al. | 4/15/2024 | Ischemic Events | Vitamin E Tocotrienol [c] |
| US11925611B2 | Use of ZT-1A and analogs thereof to prevent and/or treat neurodegenerative and neurocognitive disorders | Dandan Sun, et al. /University of Pittsburgh, US Department of Veterans Affairs VA | 3/12/2024 | Neurodegenerative symptom | Small molecule/ZT-1A Analogs [c] |
| JP2023181739A | Pharmaceutical compositions and methods for treating CADASIL patients | 敏樹 水野, et al. /Kyoto Prefectural Public Univ Corp | 12/25/2023 | Stroke symptom | Small molecule/lomerizine Hydrochloride [c] |
| TWI810499B | A composition for managing CADASIL | 家喜 袁, et al. | 8/1/2023 | General CADASIL Symptoms | Vitamin E Tocotrienol [c] |
| GB202112319D0 | Antisense oligonucleotides for the treatment of CADASIL | ProQR Therapeutics II BV | 10/13/2021 Application Ceased | General CADASIL Symptoms | Antisense-nucleotide therapy [d] |
| PH12020500376A1 | Use of cerebrolysin | Stefan Winter, et al. | 12/07/2020 | Not specified | Cerebrolysin [f] |
| US20220168325A1 | Methods to promote cerebral blood flow in the brain | Mark T. Nelson, et al. / University of Vermont | 10/1/2020 | Increasing bloodflow | Increasing PIP2 through small molecule or analog [c] |
| WO2020142373A1 | Methods and materials for treating leukodystrophies | James F. Meschia | 7/9/2020 | Leukodystrophy | CGRP inhibitor via antibody or SMI [e, c] |
| ES2692363T3 | Therapeutic compositions of mRNA and its use to treat diseases and disorders | Michael Heartlein / Translate Bio Inc | 12/03/2018 | Not specified | mRNA therapy with Fc targeting components [d,e] |
| US20160324929A1 | Methods of treating diseases associated with PPARγ | Marilyn J. Cipolla/University of Vermont and State Agricultural College | 7/02/2018 - Abandoned | PPAR-γ agonists to treat cognitive disorder | Activation of PARP-γ through relaxin [f] |
| JP2017148053A | Means and methods for modulating notch3 protein expression and/or the coding region of notch3; compositions and use thereof in the treatment of cadasil | オベルステイン, et al. | 8/31/2017 | NOTCH3mutRNA | Anti-sense oligonucleotides [d] |
| RU2461569C2 | Agonist notch3 antibodies and application thereof for treating notch3-associated diseases | Кан ЛИ (US), et al. | 9/20/2012 | Ligand binding domain of NOTCH3 | Activation of NOTCH3 via antibody [e] |
| US20080312189A1 | Cadasil Treatment with Cholinesterase Inhibitors | Raymond Pratt/Eisai Co Ltd | 12/18/2008 | Not specified | Cholinesterase inhibitors [c] |

**DETECTION METHODS**

| Patent Number | Title | Author/Company | Date Published | Method |
|---|---|---|---|---|
| CN116814763A | CADASIL gene detection kit and application thereof | 陈旭, et al. /Shanghai Lenggang Biotechnology Co ltd | 9/29/2023 | PCR |
| KR20230101339A | Diagnostic marker for CADASIL and uses thereof | 최재철, et al. | 7/6/2023 | Serological analysis of NOTCH3 ECD |
| US20230167503A1 | Method and system of diagnosing and treating neurodegenerative | Robert Meller/Morehouse School of Medicine Inc | 6/1/2023 | RNA sequencing |

| | disease and seizures | | | |
|---|---|---|---|---|
| WO2023282900A1 | Method and system of diagnosing and treating neurodegenerative disease and seizures | Robert Meller, Simon P. Roger | 1/12/2023 | RNA sequencing |
| CN110042154A | A kind of detection method and its primer sets of NOTCH3 gene mutation | 裴敏燕/Shanghai Unioon Medical Laboratory Co Ltd | 7/23/2019 | PCR |
| WO2018130665A1 | Method for investigating cerebral blood flow in a subject | オベルステイン, et al. / Leids Universitair Medisch Centrum LUMC | 7/19/2018 | fMRI analysis of cerebral blood flow |
| CN106148504A | The detection primer of dominant heredity EEG source location related gene and kit and purposes | 苏敬敬, et al. / Ninth Peoples Hospital Shanghai Jiaotong University School of Medicine | 11/23/2016 | PCR |
| CN105331614A | Method, primers and kit for detecting CADASIL (Cerebral Autosomal Dominant Arteriopathy with Subcortical Infarcts and Leukoencephalopathy) pathogenic gene mutation | 王培昌, et al./ Xuanwu Hospital | 2/17/2016 | PCR |
| CN1763196A | Gene mutation type and gene order surveying method | 陈彪, et al./Xuanwu Hospital | 9/08/2010 | Sequencing/PCR |
| JP2010042012A | Gene involved in CADASIL, method of diagnosis and therapeutic application | Elisabeth Tournier-Lasserve, et al./ Assistance Publique Hopitaux de Paris APHP Institut National de la Sante et de la Recherche Medicale INSERM | 2/25/2010 | Unspecified |
| FR2751985B1 | Gene Involved in CADASIL, diagnostic method and therapeutic application | Lasserve Elisabeth Tournier, et al. / Institut National de la Sante et de la Recherche Medicale INSERM | 10/29/1999 - Expired | Unspecified |
| CA2193564A1 | Method for the indirect genotypic diagnosis of CADASIL | Anne Joutel, et al. | 6/22/1997 - Abandoned | PCR |
| **MODELS** | | | | |
| CN115820733A | Method for establishing CADASIL disease model dog | 米继东, et al. /rain Disorders Research Center Of Capital Medical University (beijing Institute For Brain Disorders) et al. | 3/21/2023 | Canine Model of CADASIL |
| CN109679922A | The preparation method of CADASIL patient-specific vascular smooth muscle cells and vascular endothelial cell | 曲静, et al. /Institute of Zoology of CAS | 4/26/2019 | Patient derived cells |

[c] - Pharmaceuticals; [d]- Nucleic Acid Therapeutics; [e]- Immunotherapies; [f]- Biologic Therapies

**Patent Analysis**

There are currently 17 patented treatment methods for CADASIL treatment as of June 2025. Symptomatic treatments and targeted immunotherapies are the current approaches being patented (**Table III.**). Of these patents, nine protect the use of pharmaceuticals, three protect the use of nucleic acid therapies, four for immunotherapies, and three are biologic treatments. Pharmaceuticals or drug-based patents aim to relieve symptoms in patients. The gene expression-based patents focus on different areas, with one protecting the much more general use of mRNA to express a therapeutic protein that would induce an immune response, such as an antibody (ES2692363T3). The other protects the right to use antisense oligonucleotides with the goal of reducing expression of the problematic proteins (JP2017148053A). There is a noticeable lack of patented cell therapies for CADASIL treatment. The five immunologic-based patents each target a different molecule or process, utilizing an immunogenic protein such as a growth factor to promote the survival of VSMCs (US20240269234A1), or an antibody to activate NOTCH3 expression (RU2461569C2). The biologic therapies feature proteins that are not antibodies, growth factors, or cytokines that would be administered to reduce disease symptoms.

Aside from CADASIL treatment, detection methods have also been patented. There are ten recent patents for CADASIL detection, six are PCR based, two are RNAseq based, one is MRI based, and one is a serological test of the NOTCH3 ECD. There are two patented models for CADASIL, the CN115820733A patent is for the development of a CADASIL disease model with dogs, while the other patent CN109679922A is for the preparation and maintenance of CADASIL patient specific cells. A limited number of models have been specifically patented for CADASIL. These findings are also summarized in **Figure III**. The patents listed in the table were found using Google Patents using the search quarry "TI=(cadasil) OR AB=(cadasil)" to identify patents with CADASIL in the title or abstract.

**Conclusion**

Although there are no proven therapeutic treatments for CADASIL, it cannot be denied that there are various pathologic hallmarks of the disease that are great targets for continued research. This review underscores the growing potential of immunotherapy, as well as cell and gene therapy, in modulating the progression of CADASIL by directly addressing the underlying genetic and molecular mechanisms. Looking at the current status of treatments for CADASIL, the content in this review promotes research and development for continued application of cell and gene therapy strategies for disease modulation and treatment. The continued application of genomic technologies and precision medicine tools will be essential in refining these approaches and overcoming the challenges inherent to targeting rare monogenic neurovascular disorders like CADASIL. Research efforts should be continued to overcome the challenges in using immunotherapy for CADASIL treatment and will build on the foundations of the clinical trials discussed in this article.

**Acknowledgement**

Figures were created using BioRender.com and are included with permission from the creator
FIG 1 Created in BioRender. Haile, S. (2025) https://BioRender.com/jbs5yjh; Fig 2 Created in BioRender. Haile, S. (2025) https://BioRender.com/i8cu4op; Fig 3 Created in BioRender. Haile, S. (2025) https://BioRender.com/jbs5yjh

**Bibliography**

1. Bousser MG, Tournier-Lasserve E. Summary of the proceedings of the First International Workshop on CADASIL. Paris, May 19-21, 1993. Stroke [Internet]. 1994 [cited 2024 Mar 28];25:704–7. Available from: https://www.ahajournals.org/doi/10.1161/01.STR.25.3.704

2. Takahashi K, Adachi K, Yoshizaki K, Kunimoto S, Kalaria RN, Watanabe A. Mutations in NOTCH3 cause the formation and retention of aggregates in the endoplasmic reticulum, leading to impaired cell proliferation. Hum Mol Genet. 2010;19:79–89.

3. Wang J, Zhang L, Wu G, Wu J, Zhou X, Chen X, et al. Correction of a CADASIL point mutation using adenine base editors in hiPSCs and blood vessel organoids. Journal of Genetics and Genomics [Internet]. 2024 [cited 2024 May 20];51:197–207. Available from: https://www.sciencedirect.com/science/article/pii/S1673852723001108

4. Joutel A, Corpechot C, Ducros A, Vahedi K, Chabriat H, Mouton P, et al. Notch3 mutations in CADASIL, a hereditary adult-onset condition causing stroke and dementia. Nature. 1996;383:707–10.

5. Dupé C, Guey S, Biard L, Dieng S, Lebenberg J, Grosset L, et al. Phenotypic variability in 446 CADASIL patients: Impact of NOTCH3 gene mutation location in addition to the effects of age, sex and vascular risk factors. J Cereb Blood Flow Metab [Internet]. 2023 [cited 2024 Mar 31];43:153–66. Available from: https://www.ncbi.nlm.nih.gov/pmc/articles/PMC9875352/

6. Kast J, Hanecker P, Beaufort N, Giese A, Joutel A, Dichgans M, et al. Sequestration of latent TGF-β binding protein 1 into CADASIL-related Notch3-ECD deposits. Acta Neuropathologica Communications [Internet]. 2014 [cited 2024 Apr 21];2:96. Available from: https://doi.org/10.1186/s40478-014-0096-8

7. Joutel A, Andreux F, Gaulis S, Domenga V, Cecillon M, Battail N, et al. The ectodomain of the Notch3 receptor accumulates within the cerebrovasculature of CADASIL patients. J Clin Invest [Internet]. 2000 [cited 2024 Apr 18];105:597–605. Available from: https://www.jci.org/articles/view/8047

8. Ferrante EA, Cudrici CD, Boehm M. CADASIL: new advances in basic science and clinical perspectives. Current Opinion in Hematology [Internet]. 2019 [cited 2024 Mar 28];26:193. Available from: https://journals.lww.com/co-hematology/abstract/2019/05000/cadasil__new_advances_in_basic_science_and.12.aspx

9. Opherk C, Peters N, Herzog J, Luedtke R, Dichgans M. Long-term prognosis and causes of death in CADASIL: a retrospective study in 411 patients. Brain [Internet]. 2004 [cited 2024 Apr 8];127:2533–9. Available from: https://doi.org/10.1093/brain/awh282

10. Dichgans M, Mayer M, Uttner I, Brüning R, Müller-Höcker J, Rungger G, et al. The phenotypic spectrum of CADASIL: Clinical findings in 102 cases. Annals of Neurology [Internet]. 1998 [cited 2024 Mar 28];44:731–9. Available from: https://onlinelibrary.wiley.com/doi/abs/10.1002/ana.410440506

11. Chabriat H, Joutel A, Dichgans M, Tournier-Lasserve E, Bousser M-G. Cadasil. The Lancet Neurology [Internet]. 2009 [cited 2024 Mar 25];8:643–53. Available from: https://www.proquest.com/docview/201486892/abstract/74CF3869B9784C2BPQ/1

12. Hack RJ, Rutten J, Lesnik Oberstein SA. CADASIL. In: Adam MP, Feldman J, Mirzaa GM, Pagon RA, Wallace SE, Bean LJ, et al., editors. GeneReviews® [Internet]. Seattle (WA): University of Washington, Seattle; 1993 [cited 2024 Apr 3]. Available from: http://www.ncbi.nlm.nih.gov/books/NBK1500/

13. Mizuno T, Mizuta I, Watanabe-Hosomi A, Mukai M, Koizumi T. Clinical and Genetic Aspects of CADASIL. Front Aging Neurosci [Internet]. 2020 [cited 2024 Apr 7];12:91. Available from: https://www.frontiersin.org/article/10.3389/fnagi.2020.00091/full

14. Joutel A, Vahedi K, Corpechot C, Troesch A, Chabriat H, Vayssière C, et al. Strong clustering and stereotyped nature of Notch3 mutations in CADASIL patients. The Lancet [Internet]. 1997 [cited 2024 Dec 19];350:1511–5. Available from: https://www.thelancet.com/journals/lancet/article/PIIS0140-6736(97)08083-5/abstract

15. Ospina C, Arboleda-Velasquez JF, Aguirre-Acevedo DC, Zuluaga-Castaño Y, Velilla L, Garcia GP, et al. Genetic and nongenetic factors associated with CADASIL: A retrospective cohort study. Journal of the Neurological Sciences [Internet]. 2020 [cited 2024 Apr 7];419:117178. Available from: https://linkinghub.elsevier.com/retrieve/pii/S0022510X20305141

16. Jia X, Ling C, Li Y, Zhang J, Li Z, Jia X, et al. Sex differences in frontotemporal atrophy in CADASIL revealed by 7-Tesla MRI. NeuroImage: Clinical [Internet]. 2023 [cited 2024 Apr 7];37:103298. Available from: https://linkinghub.elsevier.com/retrieve/pii/S2213158222003631

17. Wang T, Baron M, Trump D. An overview of Notch3 function in vascular smooth muscle cells. Progress in Biophysics and Molecular Biology [Internet]. 2008 [cited 2024 Apr 7];96:499–509. Available from: https://linkinghub.elsevier.com/retrieve/pii/S0079610707000508

18. Ling C, Liu Z, Song M, Zhang W, Wang S, Liu X, et al. Modeling CADASIL vascular pathologies with patient-derived induced pluripotent stem cells. Protein Cell [Internet]. 2019 [cited 2025 Apr 5];10:249–71. Available from: https://www.ncbi.nlm.nih.gov/pmc/articles/PMC6418078/

19. Yuan L, Chen X, Jankovic J, Deng H. CADASIL: A *NOTCH3*-associated cerebral small vessel disease. Journal of Advanced Research [Internet]. 2024 [cited 2024 Mar 28]; Available from: https://www.sciencedirect.com/science/article/pii/S2090123224000018

20. Monet-Leprêtre M, Haddad I, Baron-Menguy C, Fouillot-Panchal M, Riani M, Domenga-Denier V, et al. Abnormal recruitment of extracellular matrix proteins by excess Notch3ECD: a new pathomechanism in CADASIL. Brain [Internet]. 2013 [cited 2024 Mar 31];136:1830–45. Available from: https://www.ncbi.nlm.nih.gov/pmc/articles/PMC3673461/

21. Zellner A, Scharrer E, Arzberger T, Oka C, Domenga-Denier V, Joutel A, et al. CADASIL brain vessels show a HTRA1 loss-of-function profile. Acta Neuropathol. 2018;136:111–25.

22. Mizuta I, Nakao-Azuma Y, Yoshida H, Yamaguchi M, Mizuno T. Progress to Clarify How NOTCH3 Mutations Lead to CADASIL, a Hereditary Cerebral Small Vessel Disease.

Biomolecules [Internet]. 2024 [cited 2024 Mar 31];14:127. Available from: https://www.ncbi.nlm.nih.gov/pmc/articles/PMC10813265/

23. Tikka S, Baumann M, Siitonen M, Pasanen P, Pöyhönen M, Myllykangas L, et al. CADASIL and CARASIL. Brain Pathol [Internet]. 2014 [cited 2024 Mar 31];24:525–44. Available from: https://www.ncbi.nlm.nih.gov/pmc/articles/PMC8029192/

24. Ishiko A, Shimizu A, Nagata E, Takahashi K, Tabira T, Suzuki N. Notch3 ectodomain is a major component of granular osmiophilic material (GOM) in CADASIL. Acta Neuropathol [Internet]. 2006 [cited 2024 Apr 3];112:333–9. Available from: https://doi.org/10.1007/s00401-006-0116-2

25. Baudrimont M, Dubas F, Joutel A, Tournier-Lasserve E, Bousser M-G. Autosomal dominant leukoencephalopathy and subcortical ischemic stroke. A clinicopathological study. Stroke [Internet]. 1993 [cited 2024 Mar 31];24:122–5. Available from: https://www.ahajournals.org/doi/epdf/10.1161/01.STR.24.1.122

26. Mayer M, Straube A, Bruening R, Uttner I, Pongratz D, Gasser T, et al. Muscle and skin biopsies are a sensitive diagnostic tool in the diagnosis of CADASIL. J Neurol [Internet]. 1999 [cited 2024 Mar 31];246:526–32. Available from: https://doi.org/10.1007/s004150050398

27. Dupré N, Gueniot F, Domenga-Denier V, Dubosclard V, Nilles C, Hill-Eubanks D, et al. Protein aggregates containing wild-type and mutant NOTCH3 are major drivers of arterial pathology in CADASIL [Internet]. American Society for Clinical Investigation; 2024 [cited 2024 Mar 31]. Available from: https://www.jci.org/articles/view/175789/pdf

28. Lee SJ, Zhang X, Wu E, Sukpraphrute R, Sukpraphrute C, Ye A, et al. Structural changes in NOTCH3 induced by CADASIL mutations: Role of cysteine and non-cysteine alterations. J Biol Chem [Internet]. 2023 [cited 2024 Mar 31];299:104838. Available from: https://www.ncbi.nlm.nih.gov/pmc/articles/PMC10318516/

29. Tikka S, Peng Ng Y, Di Maio G, Mykkänen K, Siitonen M, Lepikhova T, et al. CADASIL mutations and shRNA silencing of NOTCH3 affect actin organization in cultured vascular smooth muscle cells. J Cereb Blood Flow Metab [Internet]. 2012 [cited 2024 Mar 31];32:2171–80. Available from: https://www.ncbi.nlm.nih.gov/pmc/articles/PMC3519411/

30. Rampal R, Luther KB, Haltiwanger RS. Notch signaling in normal and disease States: possible therapies related to glycosylation. Curr Mol Med. 2007;7:427–45.

31. Arboleda-Velasquez JF, Rampal R, Fung E, Darland DC, Liu M, Martinez MC, et al. CADASIL mutations impair Notch3 glycosylation by Fringe. Hum Mol Genet. 2005;14:1631–9.

32. Pan AP, Potter T, Bako A, Tannous J, Seshadri S, McCullough LD, et al. Lifelong cerebrovascular disease burden among CADASIL patients: analysis from a global health research network. Front Neurol [Internet]. 2023 [cited 2024 Apr 21];14:1203985. Available from: https://www.ncbi.nlm.nih.gov/pmc/articles/PMC10375407/

33. Koohi F, Harshfield EL, Shatunov A, Markus HS. Does Thrombosis Play a Causal Role in Lacunar Stroke and Cerebral Small Vessel Disease? Stroke [Internet]. 2024 [cited 2025 Apr 16];55:934–42. Available from: https://www.ahajournals.org/doi/10.1161/STROKEAHA.123.044937

34. Pescini F, Torricelli S, Squitieri M, Giacomucci G, Poggesi A, Puca E, et al. Intravenous thrombolysis in CADASIL: report of two cases and a systematic review. Neurol Sci [Internet]. 2023 [cited 2024 Apr 21];44:491–8. Available from: https://www.ncbi.nlm.nih.gov/pmc/articles/PMC9842556/

35. Barisano G, Montagne A, Kisler K, Schneider JA, Wardlaw JM, Zlokovic BV. Blood-brain barrier link to human cognitive impairment and Alzheimer's Disease. Nat Cardiovasc Res [Internet]. 2022 [cited 2025 Apr 16];1:108–15. Available from: https://www.ncbi.nlm.nih.gov/pmc/articles/PMC9017393/

36. Mason BN, Russo AF. Vascular Contributions to Migraine: Time to Revisit? Front Cell Neurosci [Internet]. 2018 [cited 2024 Apr 21];12. Available from: https://www.frontiersin.org/articles/10.3389/fncel.2018.00233

37. Ghezali L, Capone C, Baron-Menguy C, Ratelade J, Christensen S, Østergaard Pedersen L, et al. Notch3ECD immunotherapy improves cerebrovascular responses in CADASIL mice. Annals of Neurology [Internet]. 2018 [cited 2024 Apr 3];84:246–59. Available from: https://onlinelibrary.wiley.com/doi/abs/10.1002/ana.25284

38. Oliveira DV, Coupland KG, Shao W, Jin S, Del Gaudio F, Wang S, et al. Active immunotherapy reduces NOTCH3 deposition in brain capillaries in a CADASIL mouse model. EMBO Mol Med [Internet]. 2022 [cited 2024 Apr 3];15:e16556. Available from: https://www.ncbi.nlm.nih.gov/pmc/articles/PMC9906330/

39. Cao CD, McCorkle JR, Kolesar JM. Beyond immunotherapy—treatment advances in cell-based therapy for ovarian cancer and associated challenges. Gynecol Pelvic Med [Internet]. 2024 [cited 2025 Jun 7];7:29. Available from: https://www.ncbi.nlm.nih.gov/pmc/articles/PMC11987047/

40. Geraghty AC, Acosta-Alvarez L, Rotiroti MC, Dutton S, O'Dea MR, Kim W, et al. Immunotherapy-related cognitive impairment after CAR T cell therapy in mice. Cell [Internet]. 2025 [cited 2025 Jun 7];0. Available from: https://www.cell.com/cell/abstract/S0092-8674(25)00391-5

41. Patil H, Bharadwaj RK, Dutta N, Subramanian R, Prasad S, Mamadapur M. CAR-T cell therapy in rheumatic diseases: a review article. Clin Rheumatol. 2025;

42. Peng J, Zhai X, Liu X, Huang Z, Wang Y, Wu P, et al. Beyond first-line therapy: efficacy and safety outcomes of continuing immunotherapy in extensive stage small cell lung cancer after PD-L1 inhibitor progression. Transl Oncol. 2025;52:102249.

43. Uwishema O, Ghezzawi M, Wojtara M, Esene IN, Obamiro K. Stem cell therapy use in patients with dementia: a systematic review. International Journal of Emergency Medicine [Internet]. 2025 [cited 2025 Jun 7];18:95. Available from: https://doi.org/10.1186/s12245-025-00876-6

44. Wang H, Zhang Y, Wan X, Li Z, Bai O. In the era of targeted therapy and immunotherapy: advances in the treatment of large B-cell lymphoma of immune-privileged sites. Front Immunol [Internet]. 2025 [cited 2025 Jun 7];16:1547377. Available from: https://www.ncbi.nlm.nih.gov/pmc/articles/PMC12023281/

45. Weller M, Rhun EL, Tsamtsouri L, Dummer R, Guckenberger M, Ribi K, et al. Immunotherapy or targeted therapy with or without stereotactic radiosurgery for patients with brain metastases from melanoma or non-small cell lung cancer – The ETOP 19-21 USZ-STRIKE study. Lung Cancer [Internet]. 2025 [cited 2025 Jun 7];199. Available from: https://www.lungcancerjournal.info/article/S0169-5002(24)00603-2/fulltext

46. Valera E, Spencer B, Masliah E. Immunotherapeutic Approaches Targeting Amyloid-β, α-Synuclein, and Tau for the Treatment of Neurodegenerative Disorders. Neurotherapeutics [Internet]. 2016 [cited 2024 Apr 18];13:179–89. Available from: https://www.ncbi.nlm.nih.gov/pmc/articles/PMC4720672/

47. Weller RO, Djuanda E, Yow H, Carare RO. Lymphatic drainage of the brain and the pathophysiology of neurological disease. Acta Neuropathologica [Internet]. 2009 [cited 2025 Apr 16];117:1–14. Available from: https://www.proquest.com/docview/211802590/abstract/BF670B676A2942B8PQ/1

48. Carare RO, Hawkes CA, Jeffrey M, Kalaria RN, Weller RO. Review: Cerebral amyloid angiopathy, prion angiopathy, CADASIL and the spectrum of protein elimination failure angiopathies (PEFA) in neurodegenerative disease with a focus on therapy. Neuropathology and Applied Neurobiology [Internet]. 2013 [cited 2024 Apr 3];39:593–611. Available from: https://onlinelibrary.wiley.com/doi/abs/10.1111/nan.12042

49. Weller RO, Hawkes CA, Kalaria RN, Werring DJ, Carare RO. White Matter Changes in Dementia: Role of Impaired Drainage of Interstitial Fluid. Brain Pathol [Internet]. 2014 [cited 2024 Apr 3];25:63–78. Available from: https://www.ncbi.nlm.nih.gov/pmc/articles/PMC8028907/

50. Machuca-Parra AI, Bigger-Allen AA, Sanchez AV, Boutabla A, Cardona-Vélez J, Amarnani D, et al. Therapeutic antibody targeting of Notch3 signaling prevents mural cell loss in CADASIL. J Exp Med [Internet]. 2017 [cited 2024 Apr 3];214:2271–82. Available from: https://www.ncbi.nlm.nih.gov/pmc/articles/PMC5551569/

51. Single cell type - NOTCH3 - The Human Protein Atlas [Internet]. [cited 2024 Apr 1]. Available from: https://www.proteinatlas.org/ENSG00000074181-NOTCH3/single+cell+type

52. Razi Soofiyani S, Baradaran B, Lotfipour F, Kazemi T, Mohammadnejad L. Gene Therapy, Early Promises, Subsequent Problems, and Recent Breakthroughs. Adv Pharm Bull [Internet]. 2013 [cited 2025 Jun 7];3:249–55. Available from: https://www.ncbi.nlm.nih.gov/pmc/articles/PMC3848228/

53. Jalil S, Keskinen T, Maldonado R, Sokka J, Trokovic R, Otonkoski T, et al. Simultaneous high-efficiency base editing and reprogramming of patient fibroblasts. Stem Cell Reports [Internet]. 2021 [cited 2024 Apr 3];16:3064–75. Available from: https://www.ncbi.nlm.nih.gov/pmc/articles/PMC8693657/

54. Takahashi K, Adachi K, Kunimoto S, Wakita H, Takeda K, Watanabe A. Potent inhibitors of amyloid β fibrillization, 4,5-dianilinophthalimide and staurosporine aglycone, enhance degradation of preformed aggregates of mutant Notch3. Biochemical and Biophysical Research Communications [Internet]. 2010 [cited 2024 Apr 3];402:54–8. Available from: https://www.sciencedirect.com/science/article/pii/S0006291X10018152

55. Reichart D, Newby GA, Wakimoto H, Lun M, Gorham JM, Curran JJ, et al. Efficient in vivo genome editing prevents hypertrophic cardiomyopathy in mice. Nat Med [Internet]. 2023 [cited 2024 Apr 28];29:412–21. Available from: https://www.nature.com/articles/s41591-022-02190-7

56. Henninger N, Mayasi Y. Nucleic Acid Therapies for Ischemic Stroke. Neurotherapeutics [Internet]. 2019 [cited 2024 Apr 3];16:299–313. Available from: https://www.ncbi.nlm.nih.gov/pmc/articles/PMC6554367/

57. Pan Y, Xin W, Wei W, Tatenhorst L, Graf I, Popa-Wagner A, et al. Knockdown of NEAT1 prevents post-stroke lipid droplet agglomeration in microglia by regulating autophagy. Cell Mol Life Sci [Internet]. 2024 [cited 2025 Jun 7];81:30. Available from: https://www.ncbi.nlm.nih.gov/pmc/articles/PMC10784396/

58. Ekkert A, Šliachtenko A, Grigaitė J, Burnytė B, Utkus A, Jatužis D. Ischemic Stroke Genetics: What Is New and How to Apply It in Clinical Practice? Genes (Basel) [Internet]. 2021 [cited 2025 Jun 7];13:48. Available from: https://www.ncbi.nlm.nih.gov/pmc/articles/PMC8775228/

59. O'Donnell MJ, Chin SL, Rangarajan S, Xavier D, Liu L, Zhang H, et al. Global and regional effects of potentially modifiable risk factors associated with acute stroke in 32 countries (INTERSTROKE): a case-control study. The Lancet [Internet]. 2016 [cited 2025 Jun 7];388:761–75. Available from: https://www.thelancet.com/journals/lancet/article/PIIS0140-6736(16)30506-2/abstract

60. Meschia JF, Worrall BB, Elahi FM, Ross OA, Wang MM, Goldstein ED, et al. Management of Inherited CNS Small Vessel Diseases: The CADASIL Example: A Scientific Statement From the American Heart Association. Stroke [Internet]. 2023 [cited 2024 Apr 7];54:e452–64. Available from: https://www.ahajournals.org/doi/10.1161/STR.0000000000000444

61. Rutten JW, Dauwerse HG, Peters DJM, Goldfarb A, Venselaar H, Haffner C, et al. Therapeutic NOTCH3 cysteine correction in CADASIL using exon skipping: in vitro proof of concept. Brain [Internet]. 2016 [cited 2024 Apr 3];139:1123–35. Available from: https://doi.org/10.1093/brain/aww011

62. Balinski AM, Preuss CV. Cilostazol. StatPearls [Internet]. Treasure Island (FL): StatPearls Publishing; 2024 [cited 2024 May 2]. Available from: http://www.ncbi.nlm.nih.gov/books/NBK544363/

63. Plosker GL, Gauthier S. Cerebrolysin. Drugs Aging [Internet]. 2009 [cited 2024 May 2];26:893–915. Available from: https://doi.org/10.2165/11203320-000000000-00000